\documentclass{vgtc}                          

\ifpdf
  \pdfoutput=1\relax                   
  \usepackage{graphicx}                
  \DeclareGraphicsExtensions{.pdf,.png,.jpg,.jpeg} 
\else
  \usepackage{graphicx}                
  \DeclareGraphicsExtensions{.eps}     
\fi%

\graphicspath{{figures/}{pictures/}{images/}{./}} 

\usepackage{microtype}                 
\PassOptionsToPackage{warn}{textcomp}  
\usepackage{textcomp}                  
\usepackage{mathptmx}                  
\usepackage{times}                     
\usepackage{cite}                      
\usepackage{tabu}                      
\usepackage{booktabs}                  

\usepackage{array}
\newcolumntype{P}[1]{>{\centering\arraybackslash}p{#1}}

\usepackage{multirow}
\usepackage{multicol}
\usepackage{cuted}

\usepackage{tabularx}
\usepackage{dcolumn}
\usepackage{tabularht}
\usepackage{bigdelim}
\usepackage{makecell}
\usepackage{longtable}
\usepackage{hhline}
\usepackage{textcomp}
\usepackage{graphics}
\usepackage{kotex}
\usepackage{xcolor}
\usepackage{makecell}
\usepackage{colortbl}

\usepackage[hyphens]{url}
\usepackage{xurl}
\usepackage{svg}
\usepackage{amsmath}
\usepackage{graphicx}
\usepackage{subcaption}

\onlineid{4921}

\vgtccategory{Research}

\title{VRDoc: Gaze-based Interactions for VR Reading Experience}

\author{
    Geonsun Lee\thanks{e-mail: gsunlee@umd.edu}\\ %
            \scriptsize University of Maryland %
    \and Jennifer Healey\thanks{e-mail: jehealey@adobe.com}\\ %
         \scriptsize Adobe Research %
    \and Dinesh Manocha\thanks{e-mail: dmanocha@umd.edu}\\ %
         {\scriptsize University of Maryland}
     }

\teaser{
  \centering
  \includegraphics[width=0.9\linewidth]{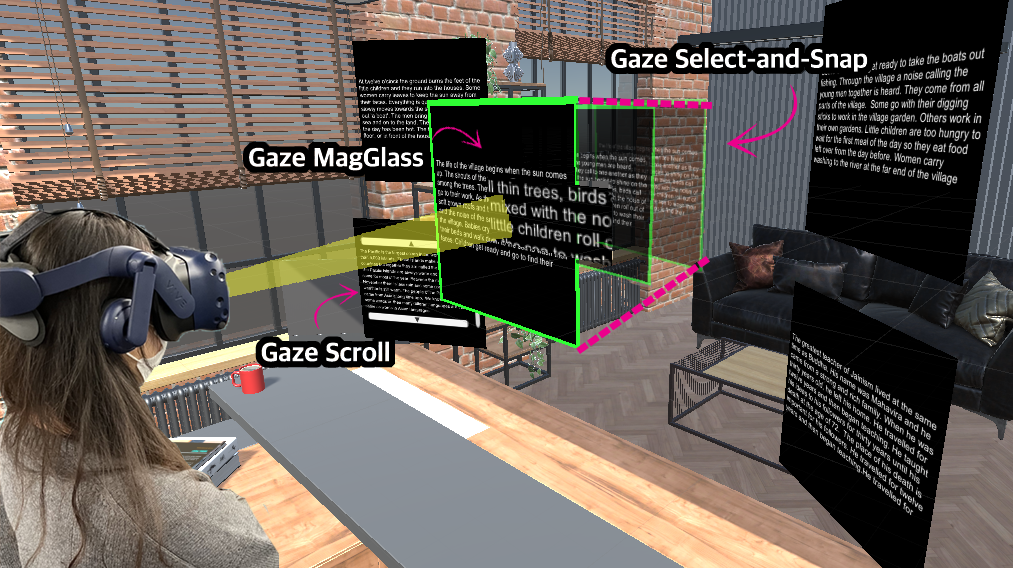}
  \caption{\textit{VRDoc} provides users a set of three gaze-based interactions that improve users' reading experience in a virtual environment; \textit{Gaze Select-and-Snap}, \textit{Gaze MagGlass}, and \textit{Gaze Scroll}. We evaluate the results and observe considerable improvement over existing interaction methods.}
  \label{fig:teaser}
}

\abstract{Virtual reality (VR) offers the promise of an infinite office and remote collaboration, however, existing interactions in VR do not strongly support one of the most essential tasks for most knowledge workers, reading. This paper presents VRDoc, a set of gaze-based interaction methods designed to improve the reading experience in VR.  We introduce three key components: Gaze Select-and-Snap for document selection, Gaze MagGlass for enhanced text legibility, and Gaze Scroll for ease of document traversal. We implemented each of these tools using a commodity VR headset with eye-tracking. In a series of user studies with 13 participants, we show that VRDoc makes VR reading both more efficient ($p < 0.01$) and less demanding ($p < 0.01$), and when given a choice, users preferred to use our tools over the current VR reading methods.%
} 

\CCScatlist{
  \CCScatTwelve{Human-centered computing}{Human computer interaction (HCI)}{Interaction paradigms}{Virtual reality};
  \CCScatTwelve{Human-centered computing}{Human computer interaction (HCI}{Interaction techniques}{};
}

\begin{document}



\maketitle

\section{Introduction}
Virtual Reality is increasingly being used for both entertainment and collaboration.  Increasingly, there is a demand for tools that can support virtual office environments \footnote{https://www.roadtovr.com/vr-apps-work-from-home-remote-office-design-review-training-education-cad-telepresence-wfh/}, including {\em Connec2, Glue, Immersed, MeetinVR, MeetingRoom, Rumii, Spatial, vSpatial, etc.}   Many of these technologies focus on the importance of remote collaboration, but overlook one of the most essential tasks of office work, reading.  Reading is essential for knowledge workers and will continue to be crucial even when co-workers want to share documents in a virtual office.   

As we look forward to a future of an immersive infinitely collaborative virtual office space, we cannot ignore the importance of a seamless reading experience being part of this virtual environment.  Currently, reading in VR is considered difficult if not impossible in a sustainable way~\cite{mirault2020using,pianzola2019virtual,rau2018speed,rzayev2018reading,wei2020reading}.  Users typically refer to many hardware-related issues with respect to the current displays which are unable to satisfy the fastidious requirements of the human visual system, including high pixel density, a large number of pixels, wide field-of-view (FOV), and a high refresh rate. At the same time, recent developments including near-eye display optics, accommodation-supporting near-eye displays, foveated displays, and vision-correcting near-eye displays~\cite{koulieris2019near} can alleviate some of these problems. It is expected that VR headsets will continue to improve in terms of display technologies and pixel resolution. 
It is also notable that, in addition to displays, other technologies such as eye tracking are being integrated as a standard feature in forthcoming HMDs~\cite{playstation_eye}.  Even with expected improvements in headset hardware, software tools for facilitating tasks will be equally important to promote seamless and sustainable productivity.  From our observational study, we have identified two major areas of improvement for VR reading tools: {\em space aware features} that facilitate selecting a document from the immersive environment and positioning it for the best reading experience and {\em document interactions} that allow users to read documents without physically leaning toward the document or manually grabbing and zooming in order to read.    
 %

Our main contributions are a set of three gaze-based interaction methods that can improve the reading experiences in VR: Gaze Select-and-Snap, Gaze MagGlass, and Gaze Scroll. Our tools are general and do not make any assumptions about the underlying task. We use standard interfaces for object manipulation along with eye tracking capabilities available in current HMDs~\cite{vive}. We first conducted a formative study to evaluate current user pain points that might impact document manipulation and reading in VR using object manipulation and canvas interaction tasks (Section 3.1). Based on this study, we identified three main challenges: difficulty positioning document objects, poor readability in current headsets, and arm fatigue (``gorilla arm''~\cite{Jang17}). To overcome these issues, we use three interaction methods: Gaze Select-and-Snap, Gaze MagGlass, and Gaze Scroll (Section 3.2). All these interaction methods  are based on eye tracking and eliminate the need to hold controllers or perform repetitive arm gestures, both of which can be tiring in long-term interactions.  With Gaze Select-and-Snap, a user can simply gaze at a virtual object tagged as a document to select it and uses a single button confirmation to bring that object into a reading view.  In the reading view, Gaze MagGlass tracks the user's eye movement and locally magnifies the text the user is reading.  Finally, when the user reaches the end of a section or page, Gaze Scroll enables the document to scroll automatically enabling seamless continued reading.

We performed user studies to evaluate our interaction tools on two tasks: reading multiple short documents and reading a long document (Section 4).  We measured tool usage, reading task completion time, and reading comprehension for both scenarios.  We then performed subjective evaluations to measure usability, effectiveness, workload, and preference using SUS and Raw TLX questionnaires. We confirmed that our tools did not increase motion sickness using the SSQ instrument for VR sickness. We observed statistically significant results that indicate that VRDoc is a more usable, less demanding, and preferred interaction method for reading in VR (Section 5). In summary, our contributions include:
\vspace{-0.1em}
\begin{itemize}
\itemsep-0.1em
\item Identifying current user pain points for VR reading experiences.
\item Developing three gaze-based user interactions to improve the VR reading experience: Gaze Select-and-Snap, Gaze MagGlass, and Gaze Scroll. 
\item Evaluating these VRDoc tools on two VR document reading tasks and observing statistically significant results for reading comprehension, task completion, usability and workload, as well as readability, efficiency, and preference.
\end{itemize}
\section{Related Work}
In this section, we briefly survey prior work in gaze-based interaction, text presentations in VR systems, and reading experiences in VR, AR, and MR systems.

\subsection{Text Presentations in VR}
Investigating text presentation on electronic devices to facilitate legibility has been an important issue for decades~\cite{dillon1990effects}. Text presentations in virtual environments present different issues than 2D monitor displays due to resolution limits and the presence of a third dimension in which to place documents. Dittrich et al. proposed a set of rules for text visualizations in 3D virtual environments~\cite{ dittrich2013legibility}. Their suggestions include having texts enlarged more than those on a 2D display. Jankowski et al. integrated text with video and 3D graphics to investigate the effects of text drawing style, image polarity, and background style (whether the background is a video or 3D)~\cite{jankowski2010integrating}. The results indicated that negative presentation of texts, such as white texts on a black background, performed better in terms of accuracy than positive presentation. Also, the billboard drawing styles, and the semitransparent white and black panels, led to a faster reading time and higher accuracy. 

Recently, Dingler et al. investigated user interface designs for displaying texts for VR reading~\cite{dingler2018vr}. They were able to identify a set of parameters such as text size, convergence, and color for an optimal text presentation. The findings indicated that users preferred texts presented with a sans-serif font and a negative presentation, that is, either having white text on a black background or having black text on a white background. There have also been efforts to investigate text presentations on 3D objects with various surfaces~\cite{wei2020reading}. It was found that a text is easier to read when it is warped around a 3D object with a single axis instead of two axes. Detailed design recommendations on the field of view and text boxes were presented. While text representation is a fundamental issue for reading in VR, it should be noted that users are given the freedom to interact with texts in an infinite margin space, not a restricted 2D display. Our paper focuses on the interaction aspect to improve users' reading performance.


\begin{figure*}[h]
	\centering
	\includegraphics[width=0.9\textwidth]{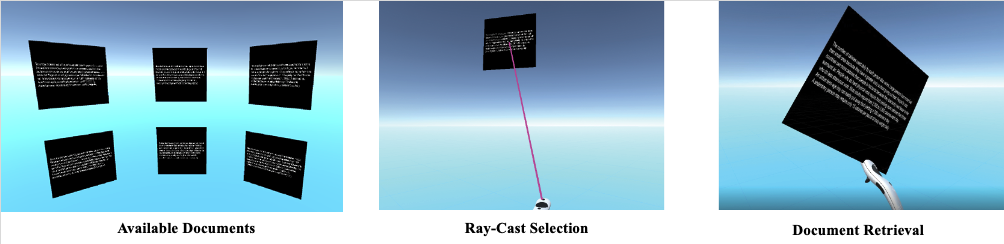}
	\vspace{-0.5em}
	\caption{In the formative study, six documents (four documents containing 100 words and two documents containing 500 words) were available to the user. Participants were able to select and manipulate the documents using raycasting and the VR controller}.
	\label{fig:formative}
	\vspace{-1.5em}
\end{figure*}

\subsection{Reading in VR/MR systems}
Reading performance depends on the devices used to display reading materials. People read slower and less accurately from computer screens than from paper~\cite{dillon1992reading, gould1987reading}, although recently the difference has been diminishing~\cite{delgado2018don}. Reading performances on tablets and paper have also been extensively compared~\cite{ chen2014comparison,connell2012effects}.
Rau et al.~\cite{rau2018speed} compared the speed and accuracy of reading in VR and AR environments with reading on an LCD monitor. The authors also compared the reading performance on two VR HMDs that differed in display quality but were otherwise similar in every way. The results indicated that users read at a slower speed with VR and AR compared with a computer screen and had a tendency to respond more accurately and faster when wearing a VR HMD of a higher pixel density.

When reading in immersive reality systems such as VR or MR, users are given more methods to interact with documents. Here, various aspects of the hardware are considered, including the presentation of the document in the VR/MR headset and the type of user input (e.g., gestures, controllers). Rzayeve et al. studied the problem of optimal text presentation type and location~\cite{rzayev2021reading}. They tested the difference in user reading experience between using the Rapid Serial Visual Presentation (RSVP) method and presenting in a paragraph. In terms of location, the study tested world-fixed, head-fixed, and edge-fixed 2D windows. They found that RSVP is effective for reading short texts when paired with edge-fixed or head-fixed locations, and a full-paragraph presentation works well with world-fixed or edge-fixed locations when minimal movement is required for the user.

For reading with MR systems, Li et al. evaluated a mixed reality experience where users read physical paper documents while seeing related artifacts such as sticky notes, figures, and videos in MR through HoloLens~\cite{li2019holodoc}.  While the study identified that readers preferred this experience to that of reading on paper, laptop, or mobile devices, this did not test or improve reading in the immersive system but showed that users enjoyed seeing ancillary material in the infinite margins surrounding documents.
Pianzola et al. investigated whether reading fiction in VR, having an immersive background, and having the ability to move your head to view texts with different orientations, affect users' absorption in the story~\cite{pianzola2019virtual}. The results show that VR enhanced users' intention to read and their affective empathy. These findings indicate that VR can be effectively exploited to promote reading in VR.

\subsection{Gaze-based Interactions}
In recent years, the interest in gaze-based interactions has surged as consumer-level eye tracking sensors have been introduced to the general public~\cite{blattgerste2018advantages}. Eye tracking technology is especially well-utilized for interactions in immersive technologies such as VR and AR~\cite{hansen2018fitts, miniotas2000application, agledahl2021magnification}. Tanriverdi and Jacob found that in a VR setting, selection with eyes showed a similar speed advantage when compared to 3D motion-tracked pointers~\cite{tanriverdi2000interacting}. Piumsomboon et al. developed a set of selection methods using natural eye movements and found that such eye gaze-based interactions could improve users’ experience while maintaining performance comparable to standard interaction techniques~\cite{piumsomboon2017exploring}. While gaze offers fast pointing, its lack of precision and difficulty of selection confirmation has been challenging. To overcome this issue, researchers combined gaze input for selection and hands for manipulation~\cite{chatterjee2015gaze+, pfeuffer2014gaze,stellmach2011designing, velloso2015empirical}. In this context, Pfeuffer et al. proposed a novel method, Gaze-touch, in which users use multi-touch gestures on interactive surfaces to control gaze-selected targets~\cite{pfeuffer2014gaze}. 
Yu et al. further incorporated gaze and hand inputs for a full 3D object manipulation in VR~\cite{yu2021gaze}. Biener et al. combined touch-based interactions with gaze for editing presentation slides in VR~\cite{biener2022povrpoint}. With a wide variety of options for utilizing gaze for interactions, it is important to select operations for which eye-tracking can play a big role so that it efficiently supplements conventional hand- or controller-based interaction methods. 

In this study, we investigate how existing conventional methods are used when reading documents, identify the drawbacks and needs of the user, and build an enhanced interaction method designed for reading documents in a virtual environment.



\section{Reading in VR}
Reading documents in a virtual reality typically involves reading a 2D document placed in a window in a 3D virtual world. This is different from users’ experience of reading in 2D displays, as document windows can be positioned at various depths and orientations. This gap might cause discomfort and inconvenience in terms of VR reading, resulting in users being reluctant to attempt this activity on a VR platform. Since prior work has found that users did not prefer head-fixed presentation when reading a paragraph of texts in VR~\cite{rzayev2021reading} our goal was to study and develop tools that allow the reader to move freely in VR, select and attach/detach documents to a reading frame easily and enhance readability. We envisioned a scenario where readers might encounter multiple documents, both long and short, in VR and that users would need to select and deselect these documents to read them.  Whereas previous works have mostly focused on short paragraphs (approximately 100 words) with a uniform font size~\cite{rzayev2018reading} we instead chose to observe users' reading behavior with longer documents with different font sizes for structure.  Based on observations from our formative study, we identified user pain points for this task using current interaction techniques.

\subsection{Formative Study}
The goal of our formative study was to identify user pain points for reading in VR from selection to reading completion with currently available tools. We observed eight users interacting with six documents of varying lengths using available VR interaction techniques. 

\subsubsection{Selected Manipulation Method}
Users wore a state-of-the-art VR Headset and were provided with a set of interaction tools that are commonly used across VR platforms for object manipulation and 2D canvas interaction (as surveyed in Section 2). Manipulation was done using the HTC Vive controller. We first asked users to select document windows and make translation movements at distance using a "laser-pointer" raycasting method~\cite{steamvr2019, vrtoolkit2019}. When a VR controller button is pressed by the user, a laser, or a ray cast, is projected in the direction to which the user is pointing, and the first colliding object in the virtual world is selected. Users can also make translation movements by moving the controller while pressing the dedicated button. For 6DOF manipulation, we use a method where users can ``grab’' and move or rotate 2D windows, which is also a common object manipulation method~\cite{wang20116d}.  Here, when a direct collision is detected between the VR controller and a 2D document window while the assigned button is pressed, the object follows the motion of the controller. An example is depicted in Figure.~\ref{fig:formative}.
Based on these interactions, users can manipulate the 2D document windows that are at various orientations.


\begin{figure*}[t]
	\centering
	\includegraphics[width=1.0\linewidth]{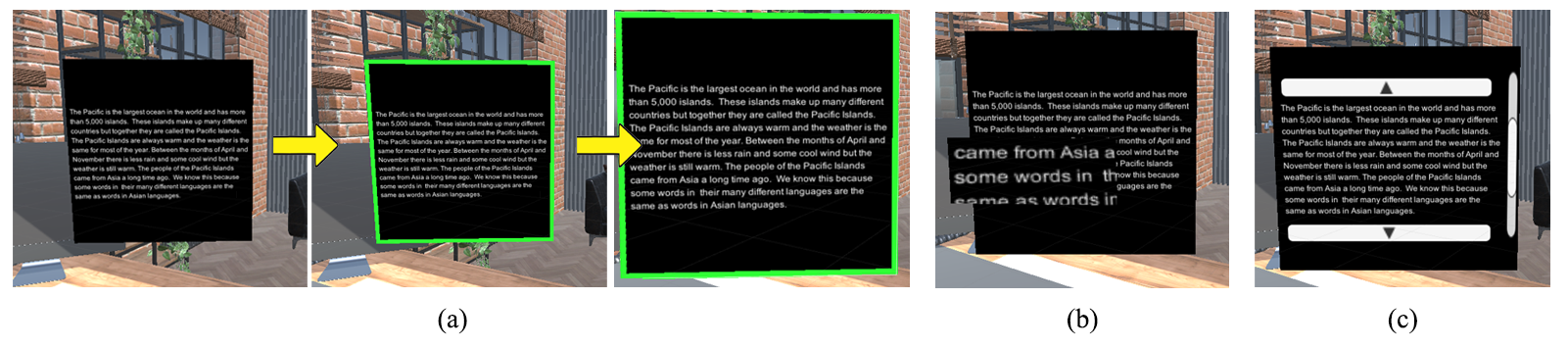}
	\caption{Interactions provided with \textit{VRDoc}. With \textit{Gaze Select-and-Snap}, (a) the document window is highlighted in green when focused and snaps in front of the user's head position when selected. \textit{Gaze MagGlass} creates a floating canvas that creates a magnifying effect shown in (b). \textit{Gaze Scroll} provides (d) gaze-activated buttons that scroll up or down by a sentence when gazed at.}
	
	\label{fig:vrdoc}
	\vspace{-1em}
\end{figure*}

\subsubsection{Participants and Procedure}
We recruited eight participants (three females, five males). Three had previous experience in VR but not in reading texts in VR. Participants wore an HTC Vive Pro Eye~\cite{vive} for the VR HMD, which has  $1440  \times 1600$ pixels per eye ($2880 \times 1600$ pixels combined), a $90$ Hz refresh rate, and a $110^{\circ}$ for the field of view. It also provides eye tracking capabilities.

Participants start from a fixed position with six documents placed in front of them as seen in Fig.~\ref{fig:formative}. The documents were placed such that they formed a semicircle around the user’s starting position. Four of the documents were short passages with about $100$ words, while the other two were long passages with about $500$ words in length. Following Dingler et al ~\cite{dingler2018vr}, the VR text presentation guideline by having a black background color and white text color for all documents. All six documents had the same canvas size, requiring the long documents to have scroll-able windows. The participants were able to scroll by selecting a document and moving their fingers vertically on the VR controller trackpad.

Each participant was given $20$ minutes to freely read all six documents without any specific order.  We followed the experience with a semi-constructed interview. The questions focused on the general experience of reading in VR, satisfaction with current interaction techniques, and desired user features.

\subsubsection{Observations and Feedback}
Through observing users’ behavior and collecting feedback through an interview, we were able to identify the following user pain points.

\paragraph{1. Positioning:} When selecting a document, it took participants multiple attempts  to re-orient the document window into the desired position. Unlike general object manipulation, participants tended to have a preferred distance and orientation (upright) for reading. Six participants noted that positioning the document to this preferred location and orientation took more time than expected. P1 commented, ``{\em It took me a while to figure out what position worked for me the best to fully view the document.}’’ All eight participants mentioned that switching their attention between multiple documents made them more aware of this inconvenience. P3 commented, ``{\em During the trial there was a moment when the document windows start to overlap as I select and position them. The pile definitely made it difficult to identify and select the documents. I wish there could be an easy way where I can quickly pick up a document that I want to read.}’’. Five participants brought up the need to automate the positioning procedure as they already knew how they wanted the document of interest to be oriented: up right in front of their head position. 

\paragraph{2. Readability:} All eight participants reported that document readability was poor due to its resolution and distortion. P2 and P5 commonly mentioned that when reading on a 2D display, they were even able to read texts sideways, but with the VR headset the text appeared blurry and this was not possible. Five participants noted that unless the document was perfectly upright, the slant created a distortion that decreased readability. We observed that participants positioned the documents with smaller font sizes closer, effectively magnifying the font, to enhance readability.   

\paragraph{3. Arm Fatigue:} During the $20$ minute trial, participants tended to hold the VR controller up constantly to interact with the documents. This behavior was consistently observed even when users were using the controller's trackpad for scrolling since the rest of the interactions, such as selecting and moving the document, required the user to hold up the physical device. Four of the participants reported arm fatigue which is a known issue ("Gorilla Arm") for gesture-based interactions~\cite{jang2017modeling} which are prevalent in VR. This is exacerbated by the additional weight of the controller when compared to watch-based gesture interactions. Such issues of fatigue would likely increase in longer VR experiences. 
\vspace{0.2em}
Based on this formative study, we developed tools specifically to address positioning, readability, and arm fatigue.

\subsection{Our Approach: VRDoc}
To address user pain points we developed three new tools: Gaze Select-and-Snap, Gaze MagGlass, and Gaze Scroll, which we collectively refer to as VRDoc, tools for better document reading in VR. This section describes our design process from user needs to potential solutions. 

Focused on the pain points of positioning, readability, and arm fatigue, our design thinking progressed as follows: 
\begin{enumerate} 
\itemsep-0.1em
\item Users do not seem to require or desire as much object manipulation freedom when reading documents.  Given the user tendency for a specific positioning with respect to documents, we should automate and simplify moving the document to a near-optimal position once selected.  Automatically positioning the document in an upright non-skewed position will enhance readability.  
\item Users manipulated documents to effectively magnify text, but this often led to documents being positioned too close for the reader to easily contextualize their place in the document.  A better solution would be to selectively magnify the current text the user is reading.
\item VR controller use and arm gestures should be minimized. While some VR controller use may still be required, in a virtual office, for longer reading tasks the user should not need to use their arms at all, enabling them to set the controller down.  We believe this will reduce fatigue and improve the overall experience.
\end{enumerate}
We believed that the novel eye tracking capability of the Vive headset could be leveraged to develop solutions for some of these issues.  Eye-tracking is becoming increasingly available in commodity HMDs~\cite{vive} and since reading naturally evokes specific eye movements, document interactions with gaze are natural and intuitive. 
In our approach,  we use the SDK provided by HTC Vive\footnote{https://developer.vive.com/documents/718/VIVE\_Pro\_Eye\_user\_guide.pdf} for eye tracking calibration and data with the Unity game engine. The headset provides eye tracking with an accuracy of  $0.5^{\circ}$ to $1.1^{\circ}$ at 120 Hz.


\subsubsection{Positioning: Gaze Select-and-Snap}
The infinite freedom of object manipulation in 3D was actually a negative factor in document positioning.  Documents were only readable in the upright position near the users' direct line of sight.  We developed \textit{Gaze Select-and-Snap} to automate the action of selecting and positioning through the user's gaze. Prior work has established that users value the ability to select 3D objects with gaze and bring these closer to the user's hand~\cite{kennedy1993simulator}, but this is the first method designed for document objects (a virtual object that is tagged as ``document'') that both rotates the 3D object into a specific position and snaps it into a fixed effective 2D perspective specifically for reading. 

To engage Gaze Select-and-Snap, the user first directs their gaze toward the 3D document object, the gaze focus is detected and the document object is highlighted with a green stroke to visualize its selection for the user.  With a single click of the trigger button, the 3D document object is brought forward towards the head position and snapped into an effective 2D position in front of the user. The window is snapped parallel to the user's head position, ensuring the window stays upright. An example of this interaction is shown in Figure.~\ref{fig:vrdoc} (a).

When multiple documents overlap, Gaze Select-and-Snap first highlights the top document, then if the top document is not selected, Gaze Select-and-Snap sequentially brings hidden documents to the forefront until the desired document is identified and selected.

\subsubsection{Readability: Gaze MagGlass}
To improve text readability once the document was in position, we incorporated a magnifying glass effect that is activated by users' eye movements: ~\textit{Gaze MagGlass}.  For low-vision computer users, video-based eye trackers have been used effectively to increase the on-screen magnification in traditional computing settings~\cite{wittich2018effectiveness, maus2020gaze}, however, to the best of our knowledge, this is the first implementation of interactive selective text magnification in VR. To enhance usability, \textit{Gaze MagGlass} is only activated when (1) the user gazes at a document, (2) the document is within a certain distance ($< 0.5 m$), and (3) when the user gazes at the document window for more than $1.5$ seconds. These robust heuristics were designed to ensure that the document in view is the specific document that the user wants to read. 

When the activation conditions are met, a second virtual camera is created on the collision point of the user’s gaze and the document object. The virtual camera is perpendicular to the document while following the user’s gaze.  The captured scene is rendered at a texture of a 2d plane that is rendered in front of the main camera. The field of view of the virtual camera and the distance from the document object are heuristically determined so that it magnifies the document by 150\% with a size that covers approximately $4$ to $5$ words of three consecutive sentences.

Directly applying raw gaze position data to the virtual camera causes great jittering as eye tracker data are inherently noisy and include tracking errors. This worsens the user experience as the jittering is visualized in a magnified way. To address this issue, we generally follow the saccade detection and smoothing algorithm from~\cite{kumar2007guide} so that the position of \textit{Gaze MagGlass} is calculated as a weighted mean of the set of points within a fixation window.

\textit{Gaze MagGlass} is automatically initiated when the activation conditions are met but can be manually turned on or off by the user if necessary. Note that activation conditions and the degree of magnification were heuristically determined for the study. An example of the activation is depicted in Figure.~\ref{fig:vrdoc} (b).

\subsubsection{Gaze Scroll}
Since longer documents are rarely considered in VR reading studies, Gaze Scroll is the first tool designed specifically to help alleviate fatigue when reading longer documents.  The objective of gaze scroll is to avoid the necessity of users having to re-engage with the controller after they have begun reading.  At this point, the document should be snapped into the 2D reading position and the user should be able to put the controller down.  To facilitate document navigation without a controller, two buttons were placed within a document each on the top and the bottom of the window, as seen in Figure~\ref{fig:vrdoc} (c). When the user's gaze reaches the button for $0.5$ seconds, the document scrolls up or down by a full sentence.  Fixating the gaze on the button can increase the number of sentences. For example, if a user stares at the lower button for 2 seconds, the document scrolls down by four sentences.  This activation condition creates a controlled advancement and minimizes focal changes, which can be frequent when scrolling is rapid or uncontrolled. 

\vspace{0.5em}
\textit{VRDoc} tools are designed to facilitate a better reading experience in VR, from automating the selection and positioning of document windows to magnifying text for readability to allowing gaze-based navigation of longer documents.

\section{Evaluation}
After implementing our tools we conducted a series of user studies to investigate how users’ reading experiences changed with VRDoc tools. We aimed to answer the following research questions through our evaluation: 
\vspace{-0.3em}
\begin{itemize}
\itemsep -0.3em
    \item Does Gaze Select-and-Snap improve document handling (positioning)?
    \item Does Gaze MagGlass improve readability?
    \item Do VRDoc tools including Gaze Scroll for navigation lessen feelings of fatigue? 
    \item Overall do VRDoc tools work together to improve efficiency and usability? 
    \item Do readers prefer having access to VRDoc tools when reading in VR?
\end{itemize}
\vspace{-0.3em}
We first evaluated each tool of VRDoc by comparing it to the basic object manipulation defined in Section 3.1 (henceforth, \textit{baseline}). Then, we conducted a study where users were given access to all VRDoc tools versus a baseline.  

\begin{figure*}[h]
	\centering
	\makebox[\textwidth][c]{\includegraphics[width=1.05\textwidth]{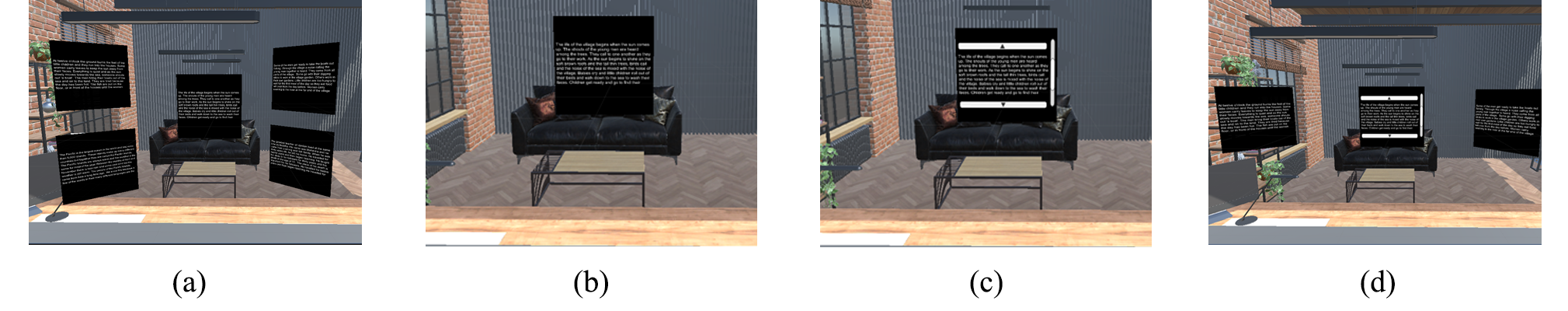}}
	\vspace{-2.0em}
	\caption{Our experiment setup for evaluation. For individual tool evaluation; (a) Task 1: Five 100-word passages, (b) Task 2: A 100-word passage, and (c) Task 3: A 500-word passage with a scroll bar. For the combined evaluation; (d) Two 100-word passages and one 500-word passage with a scroll bar.}
	\label{fig:expsetting}
	\vspace{-1.5em}
\end{figure*}

\subsection{Participants}
Thirteen participants (eight male, five female) were recruited from a convenience sample of university students for the evaluation experiment (age range $23$-$32$, $\mu=27.08$, $\sigma=2.63$). Eight participants wore glasses, three wore contact lenses, and the rest did not require vision correction. All of our participants were proficient in English. Eight of the participants had previously experienced VR systems. Participants were compensated $10$ USD after the experiment.

\subsection{Settings}
The experiments were set in a virtual office environment for immersion. The tracking area was set to $1.5m \times 1.5m$ as our task did not require much movement from the participants. We followed the guidelines suggested from previous work ~\cite{dingler2018vr} for our experiment setup. The document window contained a view box that displays a total of $9$ lines with each line comprising around $65$ characters. A white sans-serif Arial font (size $12$) was used for text and the background was set to black. The texts were left-aligned and the line spacing was set to $1.2$. The text materials were selected from the ``Asian and Pacific Speed Reading for ESL Learners’’~\cite{quinn2007asian} to guarantee a similar difficulty level. Note that the length of the text materials was slightly edited for the experiments: around 100 words for \textit{Short passages} so that they do not require scrolling and around 500 words for \textit{Long passages}.

\subsection{Experiment Procedure}
We employed within subject experiments in the order of (1) individual tool evaluation and (2) combined evaluation.
Each of the tools of VRDoc (Gaze Select-and-Snap, Gaze MagGlass, and Gaze Scroll) was compared to baseline with separate tasks. The order of tasks for individual tool evaluation and tool presentation (VRDoc, baseline) for all experiments was counterbalanced using Latin-square.

\paragraph{Individual Evaluation}
\vspace{-0.1em}
\begin{itemize}
\itemsep-0.1em
    \item \textbf{Task 1 (Gaze Select-and-Snap):} Five \textit{Short passages} are placed in front of the starting position in a semicircle, as seen in Fig.~\ref{fig:expsetting} (a). The order of the \textit{Short passages} is random. Readers are required to select each of the five documents to read them.  When the reader finishes the fifth document, a five-question reading comprehension test appears. While taking the test, readers can choose to review any of the documents by re-selecting them to help answer the questions. 
    
     \item \textbf{Task 2 (Gaze MagGlass):} One \textit{Short passage} is placed in front of the reader in the starting position as seen in Fig.~\ref{fig:expsetting}.  Gaze MagGlass tracks the reader's gaze and magnifies the font ~150\%.  When the reader reaches the end of the text, a two-question reading comprehension test appears.
     
     \item \textbf{Task 3 (Gaze Scroll):} One \textit{Long passage} is placed in front of the starting position as seen in Fig.~\ref{fig:expsetting} with no magnification. When the reader reaches the end of the text, a three-question reading comprehension test appears.
\end{itemize}

\paragraph{Combined Evaluation}
\begin{itemize}
    \item \textbf{Task 4 (VRDoc):} Two \textit{Short passages} and one \textit{Long passage} are placed in front of the starting position in a semicircle, as seen in Fig.~\ref{fig:expsetting} (a). The order of the passages is random. The reading comprehension test consists of five questions. Readers are required to select each of the three documents to read them. During the test, readers can choose to review any of the documents (by reselecting) them to help them answer the questions.
\end{itemize}

\vspace{0.2em}

Users are asked to solve a reading comprehension test after reading the given documents. 
The reading comprehension test consists of fact-check, multiple-choice questions based on the given passage. Once the participant verbally determined that they were ready to take the quiz, a 2D plane containing the questions is displayed in front of the user in the VR environment. The task completion time (TCT) was measured separately by \textit{reading time} and \textit{solving time}. The \textit{reading time} is defined as the time from the start of the trial until before reading the questions, and the \textit{solving time} is the time from after reading the questions until the participant verbally determines the final answer.

The user study was conducted in the following order: A participant was first introduced to the overview and goal of the experiment and filled out a demographic survey. Then, the participant performs each of the individual evaluation tasks in a given order, and finally the combined evaluation task. Each task consists of two trials, one performed using our proposed method and the other using baseline. The order of individual evaluation tasks and the order trials for all tasks were counterbalanced among the participants using Latin-square. Prior to each trial, participants filled out a pre-SSQ~\cite{kennedy1993simulator} survey and were given a tutorial including a five-minute training session. After each trial, participants filled out a SUS questionnaire~\cite{brooke1996sus} for system usability, Raw-TLX~\cite{hart2006nasa} for workload, post-SSQ for VR sickness, and a subjective evaluation questionnaire on a 5-point Likert scale for readability, effectiveness, and preference (See appendix \ref{custom_questions}). The experimenter reminded participants to consider the interaction aspect when answering the questions. They were given an additional 5-minute break in between trials. Note that, before each trial, we ran HTC Vive Pro Eye's eye calibration software to ensure proper gaze tracking.

After each task was performed, we conducted a semi-structured interview to collect participants’ comments and feedback. The entire experiment lasted around 90 minutes per participant. All trials performed by the participants were recorded in video files for accurate observation and evaluation. In addition to TCT for all tasks, we measured the time Gaze MagGlass was activated for Task 2 and Task 4.

\subsection{Results}
In this section, we divide the evaluation items collected during Task 1 through Task 4 into objective measures and subjective measures.
Note that the results are not directly comparable between tasks as the reading conditions vary. The visualized results can be found in appendix \ref{stat_charts}.

\subsubsection{Objective Measures} 

\paragraph{\textbf{TCT}}

We measured Task Completion Time (TCT), the sum of reading time and solving time for all four tasks:
\begin{itemize}
\itemsep-0.1em
    \item \textbf{Task 1}: Gaze Select-and-Snap allowed readers to complete both the reading task and the comprehension quiz faster than baseline.  Reading Time: Gaze Select-and-Snap ($\mu=237.73 sec$, $\sigma=21.20$), baseline ($\mu=266.62 sec$, $\sigma=34.01$); Solving Time: Gaze Select-and-Snap ($\mu=83.824 sec$, $\sigma=39.65$) baseline($\mu=107.810 sec$, $\sigma=50.91$). Both reading time and solving time revealed a significant difference between the two methods according the the Wilcoxon signed-rank test: reading time ($Z = -2.121, p = 0.034$),  solving time ($Z = -2.366, p = 0.018$), and TCT ($Z = -2.904, p = 0.004$). 
    
    \item \textbf{Task 2}: Readers read slightly faster with Gaze MagGlass ($\mu=38.11 sec$, $\sigma=13.40$) than the baseline ($\mu=41.31 sec$, $\sigma=10.73$) but the difference was not significant ($p = 0.388$). The solving time and TCT differences were also not significant (solving time: $p = 0.814$, solving TCT: $p = 0.530$). Solving Time: Gaze MagGlass ($\mu=29.41 sec$, $\sigma=10.39$), baseline ($\mu=28.91 sec$, $\sigma=6.99$); TCT: Gaze MagGlass  ($\mu=67.52 sec$, $\sigma=18.79$) baseline($\mu=70.21 sec$, $\sigma=13.39$).
    
    \item \textbf{Task 3}: Gaze Scroll allowed readers to read the longer passages slightly faster than baseline, however, Gaze Scroll had a negative impact on test solving time. Reading Time: Gaze Scroll ($\mu=157.17 sec$, $\sigma=20.41$), baseline ($\mu=164.48 sec$, $\sigma=19.67$); Solving Time: Gaze Scroll  ($\mu=92.85 sec$, $\sigma=36.64$) baseline($\mu=85.74 sec$, $\sigma=31.80$). Nevertheless, the differences were not significant which is also reflected in TCT: Gaze Scroll ($\mu=250.01 sec$, $\sigma=46.02$) baseline($\mu=250.21$, $\sigma=40.30$) ($p = $0.235).
    
     \item \textbf{Task 4}: With VRDoc, readers have significantly faster reading time and solving time than with baseline. Reading Time: VRDoc ($\mu=284.25 sec$, $\sigma=26.02$), baseline ($\mu=304.86 sec$, $\sigma=41.30$); Solving Time: VRDoc ($\mu=109.55 sec$, $\sigma=37.85$) baseline($\mu=135.78 sec$, $\sigma=40.76$); TCT:  VRDoc ($\mu=393.80$, $\sigma=50.48$) baseline($\mu=440.64 sec$, $\sigma=52.90$). Both reading time and solving time revealed a significant difference between the two methods according the the Wilcoxon signed-rank test: reading time ($Z = -2.432, p = 0.015$),  solving time ($Z = -3.101, p = 0.002$), and TCT ($Z = -2.746, p = 0.006$). 
\end{itemize}


\paragraph{\textbf{Reading Comprehension Test}} 
There were no significant differences ($p = 0.382$) in comprehension results in any of the four tasks between VRDoc and baseline according to the Wilcoxon signed-rank test.  This was expected as (1) VRDoc tools are designed to make the reading task easier and more efficient but do not explicitly aid comprehension, and (2) readers were allowed to review the passages after reading the questions. The results ensure that the participants put the same amount of effort into all tasks.

\paragraph{\textbf{Gaze MagGlass Activation Time}} 
For Task 2, Gaze MagGlass was activated in an average of 51.72\% of the reading time. With Task 4, where all of the tools are integrated, Gaze MagGlass was activated in an average of 67.18\% of the reading time.

\subsubsection{Subjective Measures} 

\paragraph{\textbf{Usability}} 
We measured the usability of the methods by calculating the SUS scores.
\vspace{-0.5em}
\begin{itemize}
\itemsep-0.1em
    \item \textbf{Task 1}: Readers found Gaze Select-and-Snap ($\mu=73.05$, $\sigma=13.05$) more usable than baseline ($\mu=51.67$, $\sigma=9.20$) with a significant difference according the the Wilcoxon signed-rank test ($Z = -2.905, p = 0.004$).
    \item \textbf{Task 2}: Readers found Gaze MagGlass ($\mu=58.11 $, $\sigma=11.49$) more usable  than baseline ($\mu=50.69$, $\sigma=12.25$) with a significant difference ($Z = -2.045, p = 0.041$).
    \item \textbf{Task 3}: The usability of Gaze Scroll ($\mu=60.89$, $\sigma=11.89$) scored and the baseline ($\mu=60.02$, $\sigma=8.48$), trackpad scrolling, were similar in terms of usability, showing no significant differences despite Gaze Scroll having a slightly higher score($p = 0.875$).
    \item \textbf{Task 4}: Readers found VRDoc ($\mu=73.20$, $\sigma=13.21$) more usable than baseline ($\mu=45.61$, $\sigma=14.53$) with a significant difference according the the Wilcoxon signed-rank test ($Z = -3.059, p = 0.002$).
\end{itemize}

\paragraph{\textbf{Workload}} 
For all four tasks, our proposed method had significantly lower RTLX scores than the baseline according to Wilcoxon signed-rank test.
\vspace{-0.5em}
\begin{itemize}
\itemsep-0.1em
    \item \textbf{Task 1}: Gaze Select-and-Snap ($\mu=31.35$, $\sigma=12.79$) scored lower than baseline ($\mu=51.67$, $\sigma=9.20$) with a significant difference ($Z = -2.591, p = 0.010$).
    \item \textbf{Task 2}: Gaze MagGlass ($\mu=28.89$, $\sigma=15.05$) scored lower than baseline ($\mu=33.98$, $\sigma=16.12$) with a significant difference ($Z = -2.367, p = 0.018$).
    \item \textbf{Task 3}: Gaze Scroll ($\mu=34.84$, $\sigma=13.63$) scored lower than baseline ($\mu=43.81$, $\sigma=16.49$) with a significant difference ($Z = -2.121, p = 0.034$).
    \item \textbf{Task 4}: VRDoc ($\mu=33.61$, $\sigma=12.66$) scored lower than baseline ($\mu=46.61$, $\sigma=17.65$) with a significant difference ($Z = -2.667, p = 0.008$).
\end{itemize}

\paragraph{\textbf{Readability}} 
For all four tasks, participants reported higher readability with our proposed method than with baseline. All differences were revealed to be significant by the Wilcoxon signed-rank test.
\vspace{-0.5em}
\begin{itemize}
\itemsep-0.1em
    \item \textbf{Task 1}: Gaze Select-and-Snap ($\mu=4.25$, $\sigma=0.75$) scored higher than baseline ($\mu=1.5$, $\sigma=0.67$) with a significant difference ($Z = -3.111, p = 0.002$).
    \item \textbf{Task 2}: Gaze MagGlass ($\mu=3.08$, $\sigma=0.79$) scored higher than baseline ($\mu=2.25$, $\sigma=0.62$) with a significant difference ($Z = -2.057, p =0.040$).
    \item \textbf{Task 3}: Gaze Scroll ($\mu=2.83$, $\sigma=0.58$) scored higher than baseline ($\mu=2$, $\sigma=0.60$) with a significant difference ($Z = -2.673, p = 0.008$).
    \item \textbf{Task 4}: VRDoc ($\mu=4.16$, $\sigma=1.02$) scored higher than baseline ($\mu=1.67$, $\sigma=0.78$) with a significant difference ($Z = -2.929, p = 0.003$).
\end{itemize}

\paragraph{\textbf{Effectiveness} }
We measured how `effective' readers found the methods to be on a 5-point Likert scale.
\vspace{-0.5em}
\begin{itemize}
\itemsep-0.1em
    \item \textbf{Task 1}: Readers found Gaze Select-and-Snap ($\mu=4.16$, $\sigma=0.83$) more effective than baseline ($\mu=1.58$, $\sigma=0.67$) with a significant difference ($Z = -3.097, p = 0.002$).
    \item \textbf{Task 2}: Readers found Gaze MagGlass ($\mu=2.67$, $\sigma=0.98$) more effective than baseline ($\mu=1.83$, $\sigma=0.93$) with a significant difference ($Z = -2.066, p = 0.039$).
    \item \textbf{Task 3}: The perceived effectiveness of Gaze Scroll ($\mu=3.58$, $\sigma=0.75$) was similar to that of baseline ($\mu=3.5$, $\sigma=0.67$) showing no significant statistical difference ($p = 0.763$).
    \item \textbf{Task 4}: Readers found VRDoc ($\mu=4.41$, $\sigma=0.79$) more effective than baseline ($\mu=1.83$, $\sigma=0.72$) with a significant difference ($Z = -2.965, p = 0.003$).
\end{itemize}

\paragraph{\textbf{Preference}} 
Subjects' preference for each method was measured through a 5-point Likert scale.
\vspace{-0.5em}
\begin{itemize}
\itemsep-0.1em
    \item \textbf{Task 1}: Readers preferred Gaze Select-and-Snap ($\mu=4.08$, $\sigma=0.67$) more than baseline ($\mu=1.83$, $\sigma=0.83$) with a significant difference ($Z = -2.971, p = 0.003$).
    \item \textbf{Task 2}: Readers preferred Gaze MagGlass ($\mu=2.67$, $\sigma=0.89$) more than baseline ($\mu=1.83$, $\sigma=0.58$) with a significant difference ($Z = -2.157, p = 0.031$).
    \item \textbf{Task 3}: Gaze Scroll ($\mu=3.18$, $\sigma=0.79$) had a similar preference score with baseline ($\mu=3.25$, $\sigma=0.75$) with no significant difference ($p = 0.603$).
    \item \textbf{Task 4}: Readers preferred VRDoc ($\mu=4.06$, $\sigma=0.74$) more than baseline ($\mu=2.29$, $\sigma=0.77$) with a significant difference ($Z = -3.126, p = 0.002$).
\end{itemize}

\paragraph{\textbf{Sickness}} 
We evaluated the sickness between the pre- and post-SSQs for each task. The results are shown in Table~\ref{table:exp_ssq}. The Wilcoxon signed-rank test revealed that there were no significant differences between the pre- and post-SSQ scores for the two methods in any of the tasks ($p > 0.05$).
\vspace{-0.5em}
\begin{table}[h]
	\small
	\centering
	\begin{tabular}{ >{\centering\arraybackslash}m{0.3cm} >{\centering\arraybackslash}m{1.2cm} |>{\centering\arraybackslash}m{0.5cm}>{\centering\arraybackslash}m{0.5cm}|>{\centering\arraybackslash}m{0.5cm}>{\centering\arraybackslash}m{0.5cm}|>{\centering\arraybackslash}m{0.6cm}|>{\centering\arraybackslash}m{1.0cm}}
		\specialrule{1.5pt}{0pt}{0pt}
		\multicolumn{2}{c|}{\multirow{2}{*}{\textbf{\rule{0pt}{4ex} \makecell{methods}}}} & \multicolumn{2}{c|}{pre-SSQ} & \multicolumn{2}{c|}{post-SSQ} &  \multirow{2}{*}{\rule{0pt}{4ex} Z} & \multirow{2}{*}{\rule{0pt}{4ex} \emph{p}}  \\ \hhline{~~----~}
		&& \thead{$ \mu$} &  \thead{$\sigma$} & \thead{$ \mu$} &  \thead{$\sigma$}&  & \\ \hline\hline
		\multirow{2}{*}{\textbf{\makecell{T1}}} & \emph{baseline} & 0.91 & 1.07 & 0.92 & 1.06 & -1.41 & 0.16  \\
		&\emph{Gaze Select-and-Snap} & 0.78 & 1.00 & 0.78 & 1.06 & -1.89 & 0.59 \\
		\hline
		\multirow{2}{*}{\textbf{\makecell{T2}}} &\textit{baseline} & 0.76 & 1.01 & 0.78 & 1.03 & -1.73 & 0.83  \\
		&\textit{Gaze MagGlass} & 0.75 & 1.08 & 0.73 & 1.05 & -1.55 &  0.61 \\
		\hline
		\multirow{2}{*}{\textbf{\makecell{T3}}} &\textit{baseline} & 0.83 & 1.12 & 0.84 & 1.03 & -1.64 & 0.65  \\
		&\textit{Gaze Scroll} & 0.80 & 1.04 & 0.83 & 1.06 & -1.21 &  0.59 \\
		\hline
		\multirow{2}{*}{\textbf{\makecell{T4}}} &\textit{baseline} & 0.86 & 1.06 & 0.97 & 1.05 & -1.50 & 0.42  \\
		&\textit{VRDoc} & 0.88 & 1.00 & 0.93 & 1.08 & -1.43 &  0.51 \\
		\specialrule{1.5pt}{0pt}{1pt}
	\end{tabular}
	\caption{SSQ analysis for the evaluation tasks, showing no significant differences between the pre- and post-SSQ results.}~\label{table:exp_ssq}
	\vspace*{-0.1in}
\end{table}

\vspace{-0.5em}

\section{Discussion and Analysis}
In this section, we analyze and discuss our findings on the quantitative results and post-experiment interviews by revisiting our research questions in Section 4.

\subsection{Gaze Select-and-Snap on Positioning}
Gaze Select-and-Snap showed the most promising and excellent results among the three VRDoc tools. We were able to observe readers significantly reducing time in selecting and reading document-of-interest out of multiple documents.
Three participants commented that they enjoyed using Gaze Select-and-Snap not only for bringing documents up front to read but to move around multiple documents quickly. P4 mentioned, ``I found it easier to scatter documents around with my gaze rather than holding up the controller and swaying my arm constantly. I could just look somewhere else to move away documents''. P5 and P11 commonly mentioned how with a larger VR space, they would find Gaze Select-and-Snap even more useful to find and orient document windows.

Although we cannot make direct comparisons between tasks since the conditions differ, when using baseline, the results in workload show that overall users’ perceived workload was higher with multiple short passages (Task 1) than with longer passages (Task 3). Implying that reading multiple documents was more demanding than reading a single long document as it involved frequently re-orienting document windows. This further strengthens the usability of Gaze Select-and-Snap, and how much positioning is important in reading documents in VR settings.

\subsection{Gaze MagGlass on Readability}
The quantitative results of the `readability' prove that Gaze MagGlass successfully enhances users' perceived readability when reading in VR.
Gaze MagGlass was the only feature that had an on/off option, since, even in a real-world scenario, we do not require magnification glasses at all times. For Task 2 and Task 4, we were able to observe that Gaze MagGlass was activated for more than half of the time. This means that this is a necessary feature for reading in VR that will likely be used. When participants were only provided with the baseline, they often moved their position to better read the passage or reached out to grab the document and manually move the document towards them.

Six participants commented that they anticipate this feature to be utilized when reading documents with a more complex structure such as paragraphs with varying fonts or multi-column documents. Five of the participants mentioned how when used with Gaze Scroll in Task 4, Gaze MagGlass helped in keeping track of where they are in the document.

\subsection{Gaze Scroll on Reducing Feelings of Fatigue}
Gaze Scroll showed promising results in the measured participants' subjective perception, yet there were some mixed reviews.
Four participants commented that Gaze Scroll was convenient in the initial reading phase since the buttons were naturally placed at the end of the reader's gaze. However, during the reading comprehension test phase, when the reader had to come back to the passage to look for information, they would prefer to use the controller since it was easier to quickly navigate to the top or bottom of the page. This implies that, for actual implementation, it would be a meaningful approach to integrate the existing trackpad-scrolling with Gaze Scroll to enhance users' reading experience.

In terms of fatigue, the SSQ results did not reveal a significant difference in fatigue when compared with the baseline. This may be because in the experiment setting, readers were not required to go through documents for hours as much as we do in our daily lives. However, Gaze Scroll is proved to be effective in reducing perceived workload as it showed significantly smaller TLX scores compared to the baseline.

\subsection{Usability, Efficiency, and Preference of VRDoc}
Subjective measures proved that individually, readers had a positive experience with each of the tools. When combined, we are able to see that all three subjective measures; effectiveness, readability, and preference were higher than the individual scores of each tool. This implies that with the three tools combined, the tools form a synergy effect.

As shown in the results, with VRDoc, users stated that they perceived improved readability for both long passages and short passages. We observe a similar conclusion with respect to how efficient the methods were in assisting users' reading experience. All these led to VRDoc having a significantly higher preference. Eight of the participants collectively mentioned how VRDoc presented a new possibility in reading in VR, which is what they have not imagined before. P13 commented, ``\textit{I used to think reading in virtual reality would be unbearable, but this opened my eyes that with the right tool, it is quite enjoyable.}’’. Similarly, P11 mentioned ``\textit{I especially liked how I felt in control of the reading space. With appropriate annotation tools, I can see reading in VR could be more useful in certain cases.}’’ Five participants pointed out that reader-specific interactions are essential in the virtual environment, and the lack of them prevents them from even attempting to read long texts in a VR setup.
Hence, for real-world applications, developers should consider providing readers in VR with a set of tools that tackle pain points to maximize users' performance.

\section{Conclusion, Limitations and Future work}
We present VRDoc, a set of three gaze-based interactions that improve users’ reading experience in VR. Through a formative study, we identified  major issues users face when reading documents with conventional object manipulation methods in a VR setting. 
We utilize eye-tracking as a solution to streamline basic interactions for users’ convenience while minimizing the need to continuously hold up a VR controller.  VRDoc consists of three gaze-based interactions that each addresses the aforementioned problems: \textit{Gaze Select-and-Snap} for selecting and positioning document windows, \textit{Gaze MagGlass} for improving readability through magnification, and \textit{Gaze Scroll} for scrolling long documents with gaze. We evaluated our method through a series of reading tasks 
and overall results indicated that the combined VRDoc tools significantly improve participants’ performance speed, usability with less workload when compared to using conventional methods. 

Our study offers insights on how gaze-based interactions could assist VR reading interfaces.  Our current approach assumes that good eye tracking capabilities are available in the VR system. Since participants’ performances are highly governed by the experiment settings, further investigation is needed to evaluate VRDoc in general applications corresponding to virtual offices or remote collaboration.
The detailed setting used for VRDoc were all empirically determined for our study. Further research should be conducted to investigate the optimal distance to place document windows for Gaze Select-and-Snap, the degree of magnification and size of canvas for Gaze MagGlass, and the placement of buttons and activation conditions for Gaze Scroll.

The experiment setting used in our approach assumed minimal movement  from the participants. It is anticipated that there will be a different set of interactions that users would need assistance with when VR locomotion is considered.  Our current study follows the guidelines on how to present texts in VR~\cite{dingler2018vr}. However, these guidelines might not always be applicable when we read documents with complex layouts or structures, such as academic papers with two columns or documents with multimedia resources. Similar to recent efforts on identifying the optimal format for reading in mobile phones~\cite{adobe2021liquidmode}, it would be worth investigating a detailed format of how the structures should be converted for optimal reading  experiences in VR. 

Going beyond simply consuming documents, it is notable that VR offers a possibility of a \emph{collaborative} workspace. For our future work, we aim to investigate collaborative interactions when multiple users read, share and discuss documents. The findings can be adapted to various multi-user applications such as virtual classrooms and virtual conferences.  As new display technologies and headsets are developed to deal with the challenging issues related to pixel density, pixel resolution, field-of-view, refresh rate and distortion, it would be useful to evaluate their impact on VR reading experiences.


\bibliographystyle{abbrv-doi}
\bibliography{vrdoc}

\begin{thebibliography}{10}
\bibitem{gould1987reading}
J.~D. Gould, L.~Alfaro, R.~Finn, B.~Haupt, and A.~Minuto.
\newblock Reading from crt displays can be as fast as reading from paper.
\newblock {\em Human factors}, 29(5):497--517, 1987.
\end{thebibliography}
\clearpage
\onecolumn
\appendix

\section{Subjective Questionnaires} \label{custom_questions}
\vspace{-10em}
\begin{table}[h]
    \centering
    \begin{tabular}{>{\centering\arraybackslash}m{2.0cm} |>
    {\centering\arraybackslash}m{12cm}}
    \specialrule{1.5pt}{0pt}{0pt}
    Measure & Questionnaire \\
    \hline\hline
      \emph{Readability} &  How was your perceived readability with the presented method?\\
      \emph{Effectiveness} & How effective did you find the presented method in performing the task? \\
      \emph{Preference} & How preferable was the presented method?\\
      \specialrule{1.5pt}{0pt}{1pt}
    \end{tabular}
    \caption{The questionnaires used to measure readability, effectiveness, and preference. The answers were made on a 5-point Likert scale.}
    \label{table:custom_questions}
    \vspace{-10em}
\end{table}

\vspace{-10em}
\section{Statistical Results} \label{stat_charts}
\vspace{-10em}
\begin{figure}[h]
	\centering
	\includegraphics[width=0.95\linewidth]{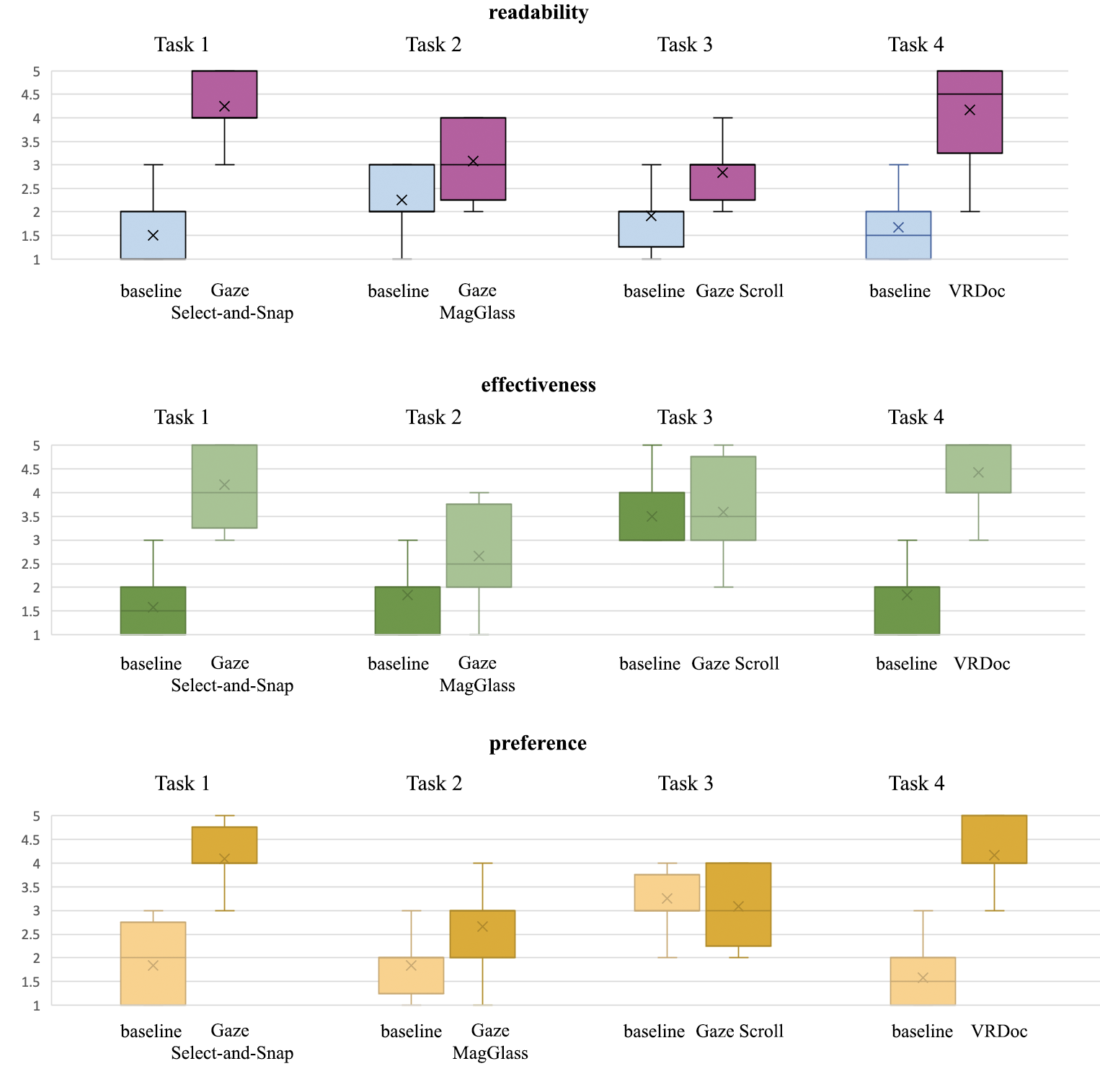}
	\caption{Visualization of the readability, effectiveness, preference results for all tasks described in Section 4.4.}	
	\label{fig:tct_stat}
	
\end{figure}

\vspace{-10em}
\begin{figure}[h]
        \small
	\centering
        \captionsetup{width=1.0\linewidth}
	\includegraphics[width=1.0\linewidth]{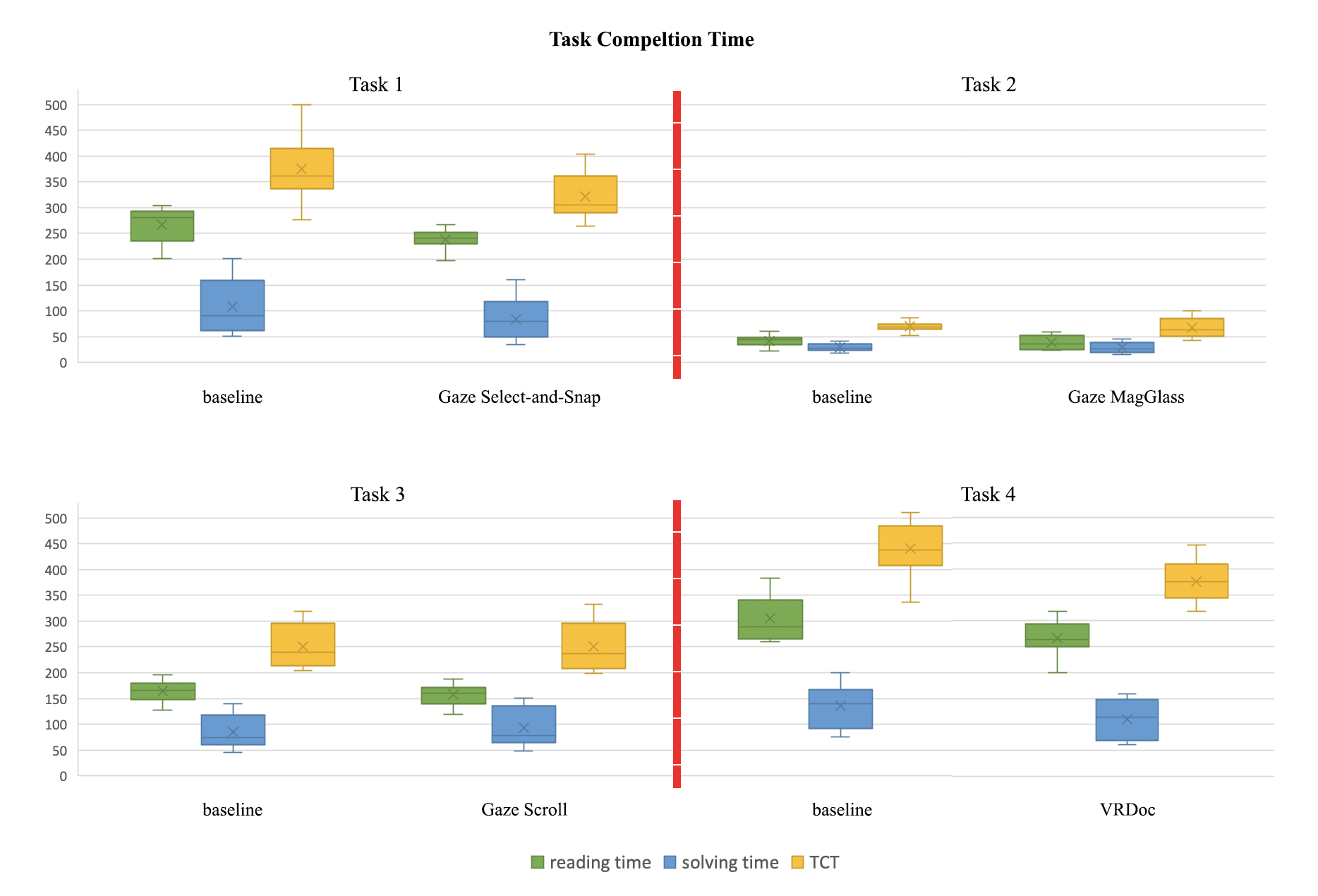}
        \vspace{-2em}
	\caption{Visualization of the TCT results for all tasks described in Section 4.4.}
        \vspace{4em}
        \includegraphics[width=1.0\linewidth]{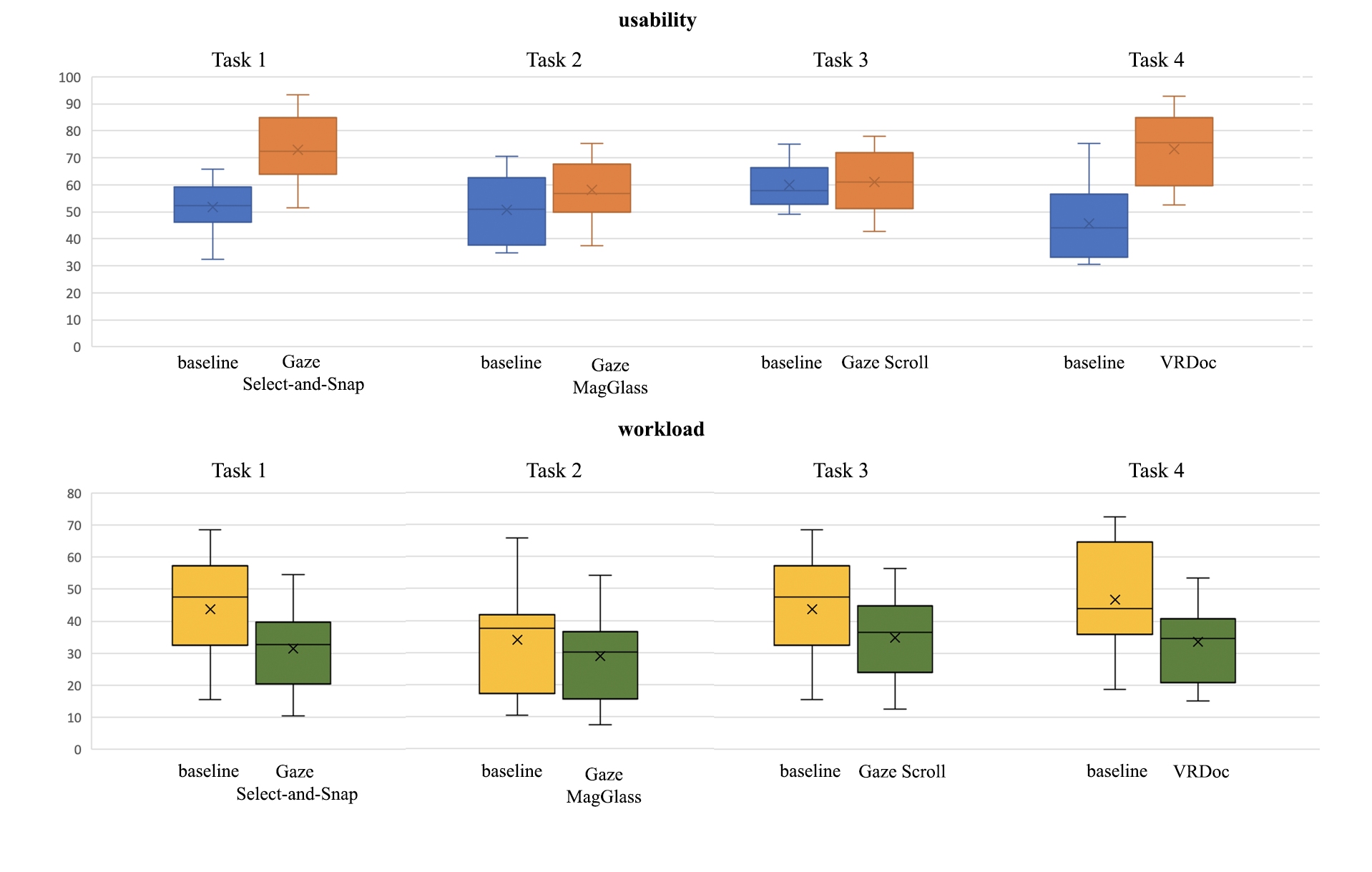}
        \vspace{-2em}
	\caption{Visualization of the SUS and RTLX results for all tasks described in Section 4.4.}	
	\label{fig:tct_stat}
	
\end{figure}

\eject
\end{document}